\documentclass[smallextended]{svjour3}       % onecolumn (second format)

\smartqed  % flush right qed marks, e.g. at end of proof
\usepackage{graphicx}
\usepackage{amssymb}
\usepackage{amsmath}
\usepackage{amsfonts}
\usepackage{mathtools}
\usepackage{cases}
\usepackage{stfloats}
\usepackage{subfigure}
\usepackage{mdwlist}
\usepackage{threeparttable}
\usepackage[all,2cell]{xy} \UseAllTwocells
\usepackage{booktabs}
\usepackage{fancyvrb}
\usepackage{rotating}
\usepackage{rotfloat}
\usepackage{pifont}
\usepackage{verbatim}
\usepackage{lineno}
\usepackage{blkarray}
\DeclarePairedDelimiter{\ceil}{\lceil}{\rceil}

\usepackage{color}

\usepackage{stfloats}
\usepackage{dsfont}
\usepackage{psfrag}
\usepackage{color,mdwlist}
\usepackage{subfigure}
\usepackage{bigstrut,array,multirow,tabularx}
\usepackage{bbm}
\DeclareMathAlphabet{\mathpzc}{OT1}{pzc}{m}{it}

\usepackage{braket}
\usepackage{dsfont}

\usepackage{graphicx}
\usepackage{amssymb}
\usepackage{amsmath}
\usepackage{cite}
\usepackage{stfloats}
\usepackage{dsfont}
\usepackage{psfrag}
\usepackage{color,mdwlist}
\usepackage{subfigure}
\usepackage{balance}
\usepackage{lipsum}
\usepackage{mathtools}
\usepackage{cuted}
\usepackage[mathscr]{euscript}
\usepackage{calrsfs}
\DeclareMathAlphabet{\pazocal}{OMS}{zplm}{m}{n}
\usepackage{bm}
\usepackage{graphicx}

\DeclareMathAlphabet{\mathsfbr}{OT1}{cmss}{m}{n}%for math sans serif (cmss)
\SetMathAlphabet{\mathsfbr}{bold}{OT1}{cmss}{bx}{n}%for math sans serif (cmss)
\DeclareRobustCommand{\msf}[1]{%
  \ifcat\noexpand#1\relax\msfgreek{#1}\else\mathsfbr{#1}\fi%for math sans serif (cmss)
}
\makeatletter
\newcommand{\msfgreek}[1]{\csname s\expandafter\@gobble\string#1\endcsname}
\makeatother

\DeclareFontEncoding{LGR}{}{} % or load \usepackage{textgreek}
\DeclareSymbolFont{sfgreek}{LGR}{cmss}{m}{n}
\SetSymbolFont{sfgreek}{bold}{LGR}{cmss}{bx}{n}
\DeclareMathSymbol{\salpha}{\mathord}{sfgreek}{`a}
\DeclareMathSymbol{\sbeta}{\mathord}{sfgreek}{`b}
\DeclareMathSymbol{\sgamma}{\mathord}{sfgreek}{`g}
\DeclareMathSymbol{\sdelta}{\mathord}{sfgreek}{`d}
\DeclareMathSymbol{\sepsilon}{\mathord}{sfgreek}{`e}
\DeclareMathSymbol{\szeta}{\mathord}{sfgreek}{`z}
\DeclareMathSymbol{\seta}{\mathord}{sfgreek}{`h}
\DeclareMathSymbol{\stheta}{\mathord}{sfgreek}{`j}
\DeclareMathSymbol{\siota}{\mathord}{sfgreek}{`i}
\DeclareMathSymbol{\skappa}{\mathord}{sfgreek}{`k}
\DeclareMathSymbol{\slambda}{\mathord}{sfgreek}{`l}
\DeclareMathSymbol{\smu}{\mathord}{sfgreek}{`m}
\DeclareMathSymbol{\snu}{\mathord}{sfgreek}{`n}
\DeclareMathSymbol{\sxi}{\mathord}{sfgreek}{`x}
\DeclareMathSymbol{\somicron}{\mathord}{sfgreek}{`o}
\DeclareMathSymbol{\spi}{\mathord}{sfgreek}{`p}
\DeclareMathSymbol{\srho}{\mathord}{sfgreek}{`r}
\DeclareMathSymbol{\ssigma}{\mathord}{sfgreek}{`s}
\DeclareMathSymbol{\stau}{\mathord}{sfgreek}{`t}
\DeclareMathSymbol{\supsilon}{\mathord}{sfgreek}{`u}
\DeclareMathSymbol{\sphi}{\mathord}{sfgreek}{`f}
\DeclareMathSymbol{\schi}{\mathord}{sfgreek}{`q}
\DeclareMathSymbol{\spsi}{\mathord}{sfgreek}{`y}
\DeclareMathSymbol{\somega}{\mathord}{sfgreek}{`w}

\DeclareMathSymbol{\svarsigma}{\mathord}{sfgreek}{`c}

\DeclareMathSymbol{\sGamma}{\mathalpha}{sfgreek}{`G}
\DeclareMathSymbol{\sDelta}{\mathalpha}{sfgreek}{`D}
\DeclareMathSymbol{\sTheta}{\mathalpha}{sfgreek}{`J}
\DeclareMathSymbol{\sLambda}{\mathalpha}{sfgreek}{`L}
\DeclareMathSymbol{\sXi}{\mathalpha}{sfgreek}{`X}
\DeclareMathSymbol{\sPi}{\mathalpha}{sfgreek}{`P}
\DeclareMathSymbol{\sSigma}{\mathalpha}{sfgreek}{`S}
\DeclareMathSymbol{\sUpsilon}{\mathalpha}{sfgreek}{`U}
\DeclareMathSymbol{\sPhi}{\mathalpha}{sfgreek}{`F}
\DeclareMathSymbol{\sPsi}{\mathalpha}{sfgreek}{`Y}
\DeclareMathSymbol{\sOmega}{\mathalpha}{sfgreek}{`W}

\DeclareRobustCommand{\mcal}[1]{%
  \ifcat\noexpand#1\relax\mathnormal{#1}\else\cal{#1}\fi
}
\DeclareRobustCommand{\BM}[1]{%
  \ifcat\noexpand#1\relax\bm{\boldUppercaseItalicGreek{#1}}\else\bm{#1}\fi
}
\makeatletter
\newcommand{\boldUppercaseItalicGreek}[1]{\csname var\expandafter\@gobble\string#1\endcsname}
\makeatother
\newcommand{\rv}[1]{\msf{#1}}

\newcommand{\M}[1]{\BM{#1}}
\let\leq\leqslant

\newcommand{\Prob}[1]{\mathbb{P}\hspace{-0.25ex}\left\{#1\right\}}

\DeclareMathVersion{timesmath}
\SetSymbolFont{letters}{timesmath}{OML}{ntxmi}{m}{it}
\DeclareMathVersion{timesmathbold}
\SetSymbolFont{letters}{timesmathbold}{OML}{ntxmi}{b}{it}

\newcommand{\slashslash}[1]{%
  \raisebox{#1}{%
    \scalebox{.7}{%
      \rotatebox[origin=c]{18}{$-$}%
    }%
  }%
}
\newcommand{\bslash}{%
  \mbox{%
   \vphantom{b}%
   \ooalign{\kern-.1em\smash{\slashslash{.9ex}}\hidewidth\cr$b$\cr}%
   \kern.05em
  }%
}
\newcommand{\slashed}[4]{%
  \mbox{%
   \vphantom{b}%
   \ooalign{\kern-#3\smash{\slashslash{#2}}\hidewidth\cr$#1$\cr}%
   \kern#4
  }}%
  
\begin{document}

\title{Information Carrier and Resource Optimization of Counterfactual Quantum Communication
}
%}
%\subtitle{Do you have a subtitle?\\ If so, write it here}

%\titlerunning{Short form of title}        % if too long for running head

\author{Fakhar Zaman         \and  Kyesan Lee \and     
        Hyundong Shin
}

%\authorrunning{Short form of author list} % if too long for running head

\institute{
		Fakhar Zaman \\
		fakhar$_-102$@khu.ac.kr \\~\\
		Kyesan Lee\\
		kyesan@khu.ac.kr\\~\\
		Hyundong Shin (Corresponding Author)\\
		hshin@khu.ac.kr \\~\\
		Department of Electronics and Information Convergence Engineering, Kyung Hee University, Yongin-si, 17104 Korea           %  \\
%           \emph{Present address:} of F. Author  %  if needed
%           \and
%           S. Author \at
%           second address
}
\date{Received: date / Accepted: date}
% The correct dates will be entered by the editor

\maketitle

%%%%%%%%%%%%%%%%%%%%%%%%%%%%%%%%%%%%%%%%%%%%%%%%%%%%%%%%%%%%%

\begin{abstract}
Counterfactual quantum communication is unique in its own way  that allows remote parties to transfer information without sending any message carrier in the channel. Although no message carrier travels in the channel at the time of successful information transmission, it is impossible to transmit information faster than the speed of light, thus without an information carrier.  In this paper, we address an important question \emph{``What carries the information in counterfactual quantum communication?''} and optimize the resource efficiency of the counterfactual quantum communication in terms of the number of channels used, time consumed to transmit 1-bit classical information between two remote parties, and the number of qubits required to accomplish the counterfactual quantum communication.

% \PACS{PACS code1 \and PACS code2 \and more}
% \subclass{MSC code1 \and MSC code2 \and more}
\end{abstract}

%%%%%%%%%%%%%%%%%%%%%%%%%%%%%%%%%%%%%%%%%%%%%%%%%%%%%%%%%%%%%%%%%%

\newpage

\section{Introduction}  \label{sec 1}
In conventional communication protocols \cite{Uetal:07:NP,SYK:14:N,Wetal:15:NP,DLL:03:PRA}, information is carried by a message carrier propagating  from sender to receiver. Unlike conventional communication, counterfactual quantum communication enables the sender (receiver) to transmit (receive) information without sending any message carrier in the channel \cite{SLAZ:13:PRL,Letal:12:PRL,ZJS:18:SR,ZSW:19:arXiv,ZSW:20:arXiv}. Elitzur and Vadmin first introduced a counterfactual phenomenon in quantum mechanics \cite{VE:93:FOP} by investigating either the absorptive object is present or not in the Mach-Zehnder interferometer---\emph{interaction free measurement} (IFM).  This invention enables one to interrogate the existence of a bomb without interrogating it with a probability $1/2$. Later, Kwiat $et~al.$ \cite{KWHZK:95:PRL} increased this probability arbitrarily close to one by cascading an array of $N$ beam splitters in series to stabilize an unstable quantum state by performing the frequent measurements---\emph{quantum Zeno effect} \cite{IHBW:90:PRA,ZJS:19:SR}. The protocol to transmit classical information in a counterfactual manner is based on a nested version of the Mach-Zehnder interferometer with $M$ outer and $N$ inner cycles---\emph{chained quantum Zeno effect}---and control the existence of an absorptive object in the interferometer in classical manner \cite{SLAZ:13:PRL,AV:19:PRA}.

In general, the number of channel usage $\eta$ and the time $T$ required to accomplish the communication task are the key parameters to determine the efficiency of  a communication system. 
In counterfactual quantum communication protocols, although the probability that the protocol is discarded approaches to zero as $M$ and $N$ increase, the performance parameters $\eta$ and $T$ increase as follows 
\begin{align}
\eta
	&
	\propto MN,\\
T
	&
	\propto\eta T_c,
\end{align} 
where $T_c$ is the time for one round trip between the sender (Alice) and receiver (Bob). 
In this paper, we optimize the resource efficiency of counterfactual quantum communication. We determine the optimal values of $M$ and $N$ to minimize $\eta$ and $T$ for the given source probability $q$ where $q=\Prob{\rv{b}=0}$ and $\rv{b}$ is 1-bit classical information Alice wants to send to Bob. In counterfactual quantum communication, although no message carrier travels in the quantum channel at the time of successful information transmission, the protocol needs a quantum channel to connect  Alice to Bob. One may ask the following questions here 
\begin{itemize}
\item
If nothing is traveling in the quantum channel at the time of successful information transmission, why do we need a quantum channel between the Alice and Bob?

\item
What carries the information in counterfactual quantum communication? 
\end{itemize} 
To answer the first question, we argue that there is a nonzero probability $\epsilon$ that the message carrier is found in the transmission channel where $\epsilon$ tends to zero as $M$ and $N$ go to infinity. In case any message carrier is found in the quantum channel, the protocol discards. In response to the second question, here we simply say the information is carried by the frequency of measurements performed in the nested interferometer. The main contribution of this paper is to address the second question and show that even if no physical particle is traveled in the transmission channel, the information carrier is transmitted from Alice to Bob followed by the resource optimization of the counterfactual quantum communication.
Currently, if the counterfactual quantum communication protocol discards in the absence of the absorptive object, the classical announcement is the only way Alice comes to know the erasure event. At the end of the paper, we demonstrate that there is no need for a classical announcement to notify Alice by making a slight modification in the original protocol and demonstrate its importance in the resource optimization of the counterfactual quantum communication. 
The rest of the paper is organized as follows. In Sec.~\ref{sec: 2}, we demonstrate how the measurement frequency acts as an information carrier in counterfactual quantum communication followed by resource optimization of counterfactual quantum communication in Sec.~\ref{sec 3}. In the end, we summarize our results and discuss the future works.

\section{Information Carrier}\label{sec: 2}

\begin{figure}[t!]
    \centering
    \subfigure[$\rv{b}=1$]
    {
        \includegraphics[width=0.7\textwidth]{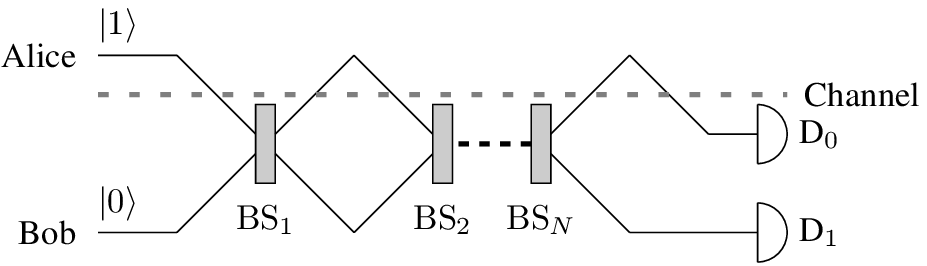}
        \label{fig:QZ:1}
    }
    \subfigure[$\rv{b}=0$]
    {
        \includegraphics[width=0.7\textwidth]{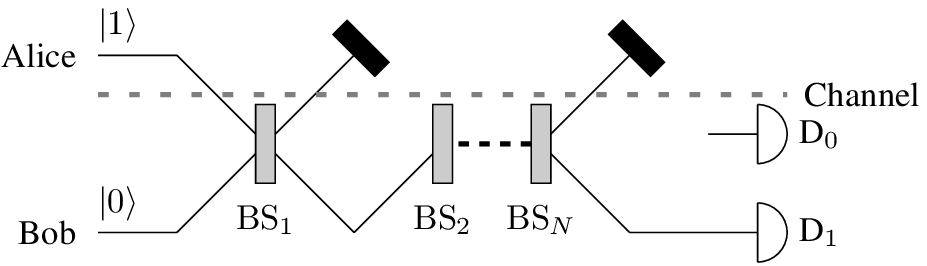}
        \label{fig:QZ:0}
    }

\caption{\emph{Semi-counterfactual quantum communication} based on the quantum Zeno effect. It is coutnerfactual only for $\rv{b}=1$. Here BS stands for unbalanced beam splitters.
}
\label{fig:QZ}
\end{figure}

\begin{figure}[t!]
    \centering
    \subfigure[$\rv{b}=1$]
    {
        \includegraphics[width=1\textwidth]{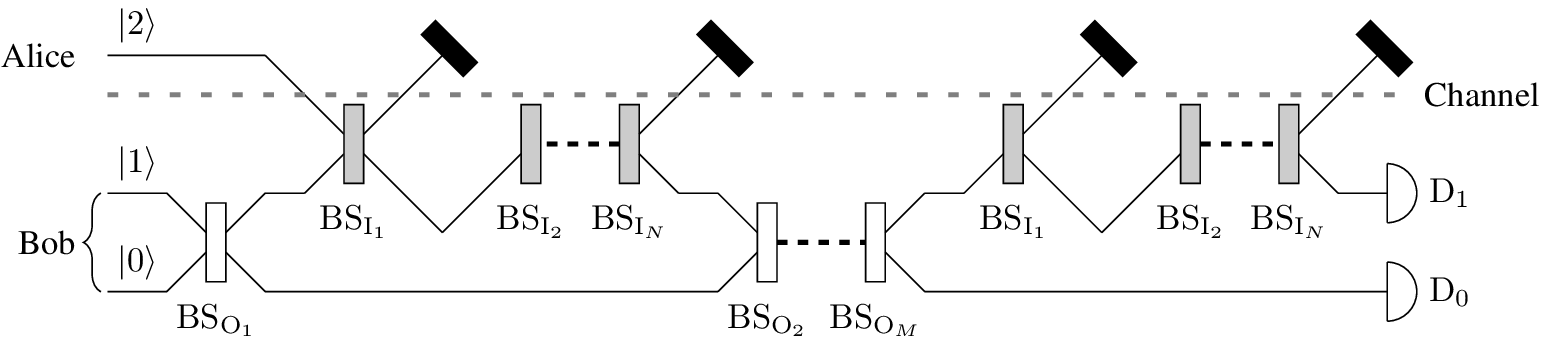}
        \label{fig:CQZ:1}
    }
    \\
    \subfigure[$\rv{b}=0$]
    {
        \includegraphics[width=1\textwidth]{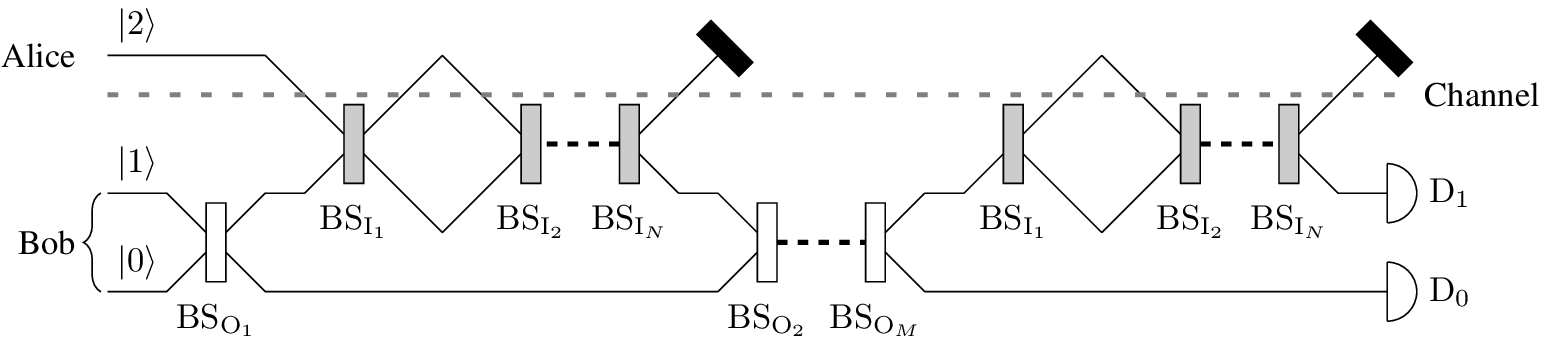}
        \label{fig:CQZ:0}
    }

\caption{\emph{Counterfactual quantum communication} based on the chained quantum Zeno effect. It is counterfactual for both classical bits $\rv{b}=0$ and $\rv{b}=1$. Here white rectangles denote $M$ outer cycles and gray rectangles represent $N$ inner cycles. 
}
\label{fig:CQZ}
\end{figure}

Counterfactual quantum communication is a unique phenomenon in quantum mechanics that allows Alice and Bob to transfer information without transmitting any physical particle over the channel. To achieve the counterfactual quantum communication, Alice maps one bit of classical information $\rv{b}$ in the presence or absence of the shutter in one arm of the interferometer. If $\rv{b}=1$, Alice introduces the shutter in her part of the interferometer to block the path of the photon as shown in Fig~\ref{fig:CQZ}. We consider that if the photon interacts with the shutter, it absorbs the photon. In case $\rv{b}=0$, Alice allows the photon to pass and reflect the component of the photon to Bob. Before going into the details of counterfactual quantum communication, we explain the working of an interferometer and show that it is \emph{semi-counterfactual} (counterfactual for $\rv{b}=1$ only). Fig.~\ref{fig:QZ} shows an array of $N$ unbalanced beam splitters (BS)  and each BS transforms the input state of the photon as 

\begin{align}
\M{U}
	=
	\begin{bmatrix}
	\cos\theta_N & -\sin\theta_N \\
	\sin\theta_N & \cos\theta_N
	\end{bmatrix},
\end{align}
where $\theta_N=\pi/\left(2N\right)$. Consider the initial state of the photon is $\ket{\psi_0}=\ket{0}$. After BS$_1$, the state of the photon is transformed as $\ket{\psi_1}=\M{U}\ket{0}=\cos\theta_N\ket{0}+\sin\theta_N\ket{1}$. If $\rv{b}=1$, Alice blocks the path of the photon with shutter. The presence of the shutter is similar to measure the state of the photon in computational basis. If the state of the photon is $\ket{1}$, the photon is absorbed by the shutter and no photon reaches at the second beam splitter. Unless the photon is absorbed by the shutter, the photon collapses back to the initial state and Bob inputs the photon to BS$_2$ (see Fig.~\ref{fig:QZ:1}). After $N$ cycles, the photon collapses to the initial state with probability $\cos^{2N}\theta_N$ which tends to one as $N\rightarrow\infty$ and D$_1$ clicks. Here it is important to note that Bob determines the presence of the shutter at Alice's side  without transmitting any physical particle over the channel.  

For $\rv{b}=0$, Alice allows the photon to pass and no measurement is performed on the photon.
The $N$ BSs transform the state of the photon to $\ket{\psi_1}=\M{U}^N\ket{0}=\ket{1}$ and D$_0$ clicks (see Fig.~\ref{fig:QZ:0}).  Bob decodes the classical bit as  $\rv{b}$ if D$_\rv{b}$ clicks and the measurement frequency $f$ is given as 
\begin{align}
f=
	\begin{cases}
		0, & \text{for~}\rv{b}=0,\\
		N, & \text{for~}\rv{b}=1.
	\end{cases}
\end{align}
However, this protocol is counterfactual for $\rv{b}=1$ only---called \emph{semi-counterfactual quantum communication}. In case $\rv{b}=0$, the photon is found in the transmission channel with certainty. 

To transfer both classical bits 0 and 1 in counterfactual way---called \emph{counterfactual quantum communication}, consider the nested version of the interferometer as shown in Fig.~\ref{fig:CQZ}.  Here white rectangles show the unbalanced beam splitters BS$_\mathrm{O}$ for outer cycles and gray rectangles denote the unbalanced beam splitters BS$_\mathrm{I}$ for inner cycles where each outer cycle has an array of $N$ BS$_\mathrm{I}$. Each BS$_\mathrm{O}$ and BS$_\mathrm{I}$  transform the state of the photon as $\M{U}_\mathrm{O}$ and $\M{U}_\mathrm{I}$, respectively where
\begin{align}
\M{U}_\mathrm{O}
	&=
	\begin{bmatrix}
	\cos\theta_M & -\sin\theta_M & 0\\
	\sin\theta_M & \cos\theta_M  & 0\\
	0			&			    & 1
	\end{bmatrix},\\
\M{U}_\mathrm{I}
	&=
	\begin{bmatrix}
	1	&	0			&	0\\
	0	& 	\cos\theta_N & -\sin\theta_N\\
	0	&	\sin\theta_N & \cos\theta_N \\
	\end{bmatrix},	
\end{align}
where $\theta_M=\pi/\left(2M\right)$. 
Assume that the quantum state of  Bob's photon can be in the superposition of $\ket{0},\ket{1}$ and $\ket{2}$ where $\ket{0},\ket{1}$ and $\ket{2}$ show the paths of the photon (see Fig.~\ref{fig:CQZ}). We consider that the initial state of the photon is $\ket{\phi_0}=\ket{0}$ and Bob inputs the photon to the first outer cycle. After BS$_{\mathrm{I}_1}$ of the first outer cycle, the initial state $\ket{\phi_0}$ transforms as follows:
\begin{align}
\ket{\phi_1}
	&=
	\M{U}_\mathrm{I}
	\M{U}_\mathrm{O}
	\ket{0}\label{eq: qutrit}\\
	&=
	\cos\theta_M\ket{0}
	+
	\sin\theta_M
	\cos\theta_N
	\ket{1}
	+
	\sin\theta_M
	\sin\theta_N
	\ket{2}.
\end{align}
If $\rv{b}=1$, Alice blocks the path of the photon with shutter (see Fig.~\ref{fig:CQZ:1}). The presence of the shutter is similar to perform the measurement on the photon. If the state of the photon is $\ket{2}$, the photon is absorbed by the shutter and no photon reaches at BS$_{\mathrm{I}_2}$. Unless the photon is absorbed by the shutter, the state of the photon collapses to\footnote{We consider $\cos\theta_N\approx 1$ for large values of $N$.} $\ket{\phi_2}=\cos\theta_M\ket{0}+\sin\theta_M\ket{1}$ and the state of the photon after BS$_{\mathrm{I}_N}$ of the first outer cycle collapses to $\ket{\phi_2}$. Bob inputs the photon to BS$_{\mathrm{O}_2}$. After $M$ outer cycles, the state of the photon is 
\begin{align}
\ket{\phi_3}
	&=
	\M{U}_\mathrm{O}^{M-1}
	\ket{\phi_2}\\
	&=
	\ket{1},
\end{align}
and detector D$_1$ clicks. Note that the measurement frequency is $f=N$ in each outer cycle for $\rv{b}=1$.

For $\rv{b}=0$, Alice allows the photon to pass and no measurement is performed on the photon.
After BS$_{\mathrm{I}_N}$ of the first outer cycle, the state of the photon transforms to 
\begin{align}
\ket{\phi_2}
	&=
	\M{U}_\mathrm{I}^N
	\M{U}_{\mathrm{O}}
	\ket{\phi_0}\\
	&=
	\cos\theta_M
	\ket{0}
	+
	\sin\theta_N
	\ket{2}.
\end{align}
Alice blocks the path of the photon independent of the value of $\rv{b}$ (see Fig.~\ref{fig:CQZ:0}). Unless the photon is absorbed by the shutter, the state of the photon collapses to the initial state $\ket{\phi_0}$. After $M$ outer cycles, the state of the photon remains in the initial state, and D$_0$ clicks. Note that the measurement frequency is $f=1$ in each outer cycle $\rv{b}=0$. At the end of the protocol, the measurement frequency $f$ is given as 
\begin{align}
f
	=\begin{cases}
	M, ~~~~&\text{for}~\rv{b}=0,\\
	MN,~~~~&\text{for}~\rv{b}=1.
	 \end{cases}\label{eq: frequency}
\end{align}
Since the measurement on the quantum state disturbs the state of the system under observation, from \eqref{eq: qutrit} and \eqref{eq: frequency}, we can see  that the measurement frequency on the qutrit itself is the information carrier in counterfactual quantum communication  and transforms the initial state to orthonormal states which are perfectly distinguishable to decode the information.

\begin{figure}[t!]	
  \centering
\includegraphics[width=0.65\textwidth]{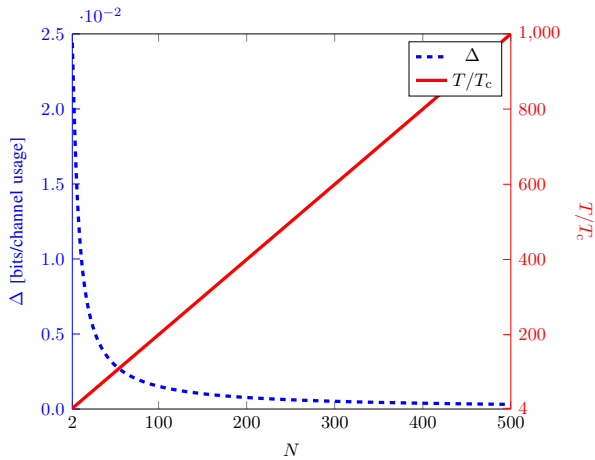}
    \caption{
      Successful transmission rate $\mathrm{\Delta}$ [bits/channel usage] and $T/T_\mathrm{c}$ for $q=1/2$ and $M_\star=\arg\max_M\mathrm{\Delta}$. As $N$ increases, the successful transmission rate $\Delta$  decreases as a logarithmic function and $T/T_\mathrm{c}$ increases linearly. }
    \label{fig:CQC q=0.5}
\end{figure}
\begin{figure}[t!]	
  \centering
\includegraphics[width=0.65\textwidth]{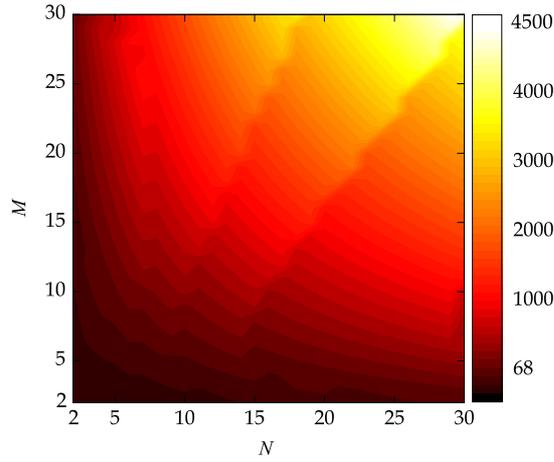}
    \caption{
      $\zeta_{M,N}$ for $q=1/2$ and $P=0.975$. Since the success probability is same for  any values of $M$ and $N$, there exist the optimal values of $M$ and $N$ such that $\lbrace M_\star,N_\star\rbrace=\arg\min\zeta_{M,N}$. For $q=1/2$, $M_\star=N_\star=2$  minimizes $\zeta_{M,N}$  as $\zeta_{\min}=68$. }
    \label{fig:CU q=0.5}
\end{figure}

\section{Resource Optimization}\label{sec 3}
In the previous section, we demonstrated that the measurement frequency carries the information in the counterfactual quantum communication (see \eqref{eq: frequency}). To ensure the full counterfactuality of the protocol, Alice and Bob use the nested interferometer with $M$ outer and $N$ inner cycles (see Fig.~\ref{fig:CQZ}) to transfer one bit of classical information without transmitting any physical particle over the channel with an average success probability $\lambda_{M,N}$ for a given source $q$ where 
\begin{align}
\lambda_{M,N}
&=
	q
	\lambda_0
	+
	\left(1-q\right)
	\lambda_1,
\end{align}
and $\lambda_0$ and $\lambda_1$ are defined as functions of $M$ and $N$ as \cite{ZJS:18:SR}
\begin{align}
\lambda_0
&=
	\cos^{2M}\theta_M,\\
\lambda_1
&=
	\prod_{m=1}^M
	\left[
		1
		-
		\sin^2\left(m\theta_M\right)
		\sin^2\theta_N
	\right]^N.
\end{align}
Although no physical particle is found in the transmission channel at the time of successful transmission, there is a nonzero probability that the photon has one round trip between Alice and Bob in each inner cycle of the nested interferometer which tends to zero as $M$ and $N$ go to infinity. Here we consider the number of channel usages $\eta$ and time $T$ required to finish the task as the primary resource of counterfactual quantum communication. Our goal is to minimize the number of channel usage and reduce the implementation time to transfer classical information counterfactually.   For a given source $q$, the resource parameters $\eta$  and $T$ to transmit 1-bit classical information counterfactually with success probability $\lambda_{M,N}$ are given as 
\begin{align}
\eta
&=
	2MN,\\
T
&=
	\dfrac{1}{2}
	\eta T_{\mathrm{c}}.
\end{align}
%%
%where $T_c$ is the time for one round trip between the Alice and Bob. 
To determine the efficiency of the counterfactual quantum communication for given $q$ corresponding to $M$ and $N$, we define the successful transmission rate $\mathrm{\Delta}$ [bits/channel usage] as 
\begin{align}
\mathrm{\Delta}
=
	\dfrac{\lambda_{M,N}}{\eta}.
\end{align}
Fig.~\ref{fig:CQC q=0.5} shows the successful transmission rate $\mathrm{\Delta}$ [bits/channel usage] and $T/T_\mathrm{c}$ as a function of $N$ corresponding to $M_\star=\arg\max_M\mathrm{\Delta}=2$ and $q=1/2$. Although $\lambda_{M,N}$ increases with $N$, we can clearly see that $\mathrm{\Delta}$ decreases as a logarithmic function of $N$, on the other hand $T/T_\mathrm{c}$ increases linearly. 
\begin{figure}[t!]	
  \centering
\includegraphics[width=0.62\textwidth]{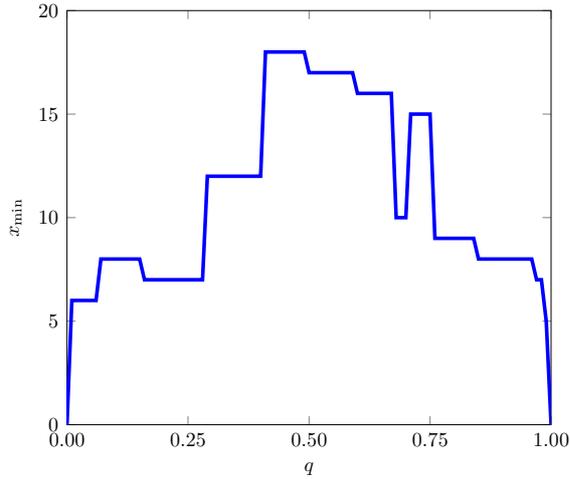}
    \caption{
     Number of transmission trials $x_{M_{\star},N_{\star}}$ as a function of $q$ for $P=0.975$. It shows the cost of achieving the same success probability with optimal values of $M$ and $N$.   }
    \label{fig:trials}
\end{figure}

To optimize the resource efficiency of the counterfactual quantum communication protocols, we fix the success probability $P$ of counterfactual quantum communication  independent of $M$ and $N$. Let $\rv{X}$ be the number of transmission trials up to the success of counterfactual quantum communication. Then $\rv{X}$ is the geometric random variable with success trial probability $\lambda_{M,N}$. Let $x_{M,N}$ be the minimum required number of transmission trials for given $P$ with $M$ outer and $N$ inner cycles. Then we have
\begin{align}
P
&
	\leq
	\Prob{\rv{X}\leq x_{M,N}}
	\\
&
	=
	\sum_{i=1}^{x_{M,N}}
	\left(1-\lambda_{M,N}\right)^{i-1}
	\lambda_{M,N}\\
&
	=
	1
	-
	\left(
		1
		-
		\lambda_{M,N}
	\right)^{x_{M,N}},\label{eq:geometric distribution}
\end{align}
leading to
\begin{align}
x_{M,N}
=
	\ceil[\Bigg]
	{
		\dfrac{\log\left(1-P\right)}{\log\left(1-\lambda_{M,N}\right)}
	}.
\end{align}
where $\lceil\cdot\rceil$ is the ceiling function. Since the success probability $P$ is independent of $M$ and $N$, we calculate  
\begin{align}
\zeta_{\min}=\min\zeta_{M,N},
\end{align}
where $\zeta_{M,N}=MNx_{M,N}$ and the performance parameters can be optimized as 
\begin{align}
\eta_{\min}
	&=
	2\zeta_{\min},\\
T_{\min}
	&=\zeta_{\min}T_\mathrm{c},\\
\mathrm{\Delta}_{\max}
	&=
	\dfrac{P}{\eta_{\min}}.	
\end{align}

For example, $\zeta_{\min}=68$ for $q=1/2$ and $P=0.975$  with $M_\star=N_\star=2$ and $x_{M_{\star}N_{\star}}=17$ as shown in Fig.~\ref{fig:CU q=0.5} where $\lbrace M_\star,N_\star\rbrace=\arg\min\zeta_{M,N}$. Here it is important to note that smaller the values of $M_{\star}$ and $N_{\star}$, larger the value of $x_{M_{\star},N_{\star}}$, which reveals that a large number of photons needs to transfer 1-bit classical information counterfactually with success probability at least $P$.  Fig.~\ref{fig:trials} shows the cost of minimizing $\eta$ and $T$.
\begin{figure}[t!]	
  \centering
\includegraphics[width=0.7\textwidth]{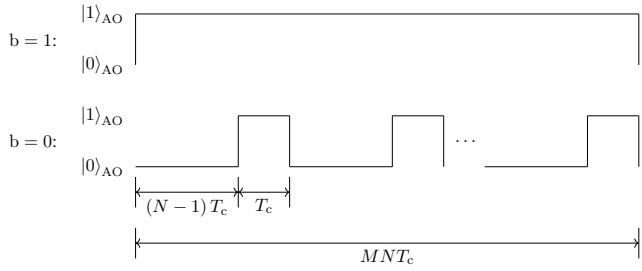}
    \caption{
     Wave functions for binary counterfactual quantum communication. Here $\ket{0\left(1\right)}_{\mathrm{AO}}$ denotes absence (presence) of the absorptive object and AO stands for the absorptive object.   }
    \label{fig:AO}
\end{figure}

In the original protocol for counterfactual quantum communication \cite{SLAZ:13:PRL}, if the photon is traveled between Alice and Bob for $\rv{b}=1$, it is absorbed by the absorptive object (shutter) and the absorptive object jumps to high energy level. It causes an erasure for both Alice and Bob. In case $\rv{b}=0$, if the photon is found in the transmission channel, it ends up at detector D$_3$ (see Fig.~1 in \cite{SLAZ:13:PRL}). It causes an erasure for Bob only. Alice has no information either the information is transferred or the protocol is discarded. To perform the resource optimization, it is important that Alice knows when the first success occurs so that she can start the transmission of the next bit in the string. To accomplish this task, Fig.~\ref{fig:AO} shows two wave functions for binary counterfactual quantum communication.
For $\rv{b}=1$, the wave function is the same as the presence of the absorptive object in the interferometer for $M$ outer and $N$ inner cycles.  In case $\rv{b}=0$, Alice transmits the wave function such that the absorptive object (shutter) is present in the last inner cycle of each outer cycle independent of $\rv{b}$. With this wave function if the photon is found in the transmission channel for $\rv{b}=0$, it causes an erasure for both Alice and Bob. This slight modification at Alice's side enables Alice and Bob to transmit one bit of classical information under the most optimal resource efficiency.

\section{Conclusion}\label{sec 4}

In this paper, we analyzed the resource efficiency of the counterfactual quantum communication and demonstrated that small values of $M$ and $N$ give better efficiency with multiple rounds of the counterfactual quantum  communication protocol ($x_{M,N}$) to increase the success probability $P$. It is already well known that the stability of the interferometer is questioned for large values of $M$ and $N$. In addition, the effect of channel noise \cite{LLZZ:18:OP,LLZZ:17:SR} and photon loss probability also increase as $M$ and $N$ increase. We showed that $M=N=2$ are the optimal values for $q=1/2$ and $P=0.975$ at the cost of $x_{M,N}=17$. 
We also demonstrated that the measurement frequency on the qutrit state acts as the information carrier in the counterfactual quantum communication. 

%%%%%%%%%%%%%%%%%%%%%%%%%%%%%%%%%%%%%%%%%%%%%%%%%%%%%%%%%%%%%

\begin{acknowledgements}
This work was supported by the National Research Foundation of Korea (NRF) grant funded by the Korea government (MSIT) (No. 2019R1A2C2007037).
\end{acknowledgements}

%%%%%%%%%%%%%%%%%%%%%%%%%%%%%%%%%%%%%%%%%%%%%%%%%%%%%%%%%%%%%%%%

%\bibliographystyle{spbasic}      % basic style, author-year citations
%\bibliographystyle{spmpsci}      % mathematics and physical sciences
%\bibliographystyle{splncs}		% Springer style for unsrt
%\bibliography{ROCQC}

\end{document}